\documentclass[twocolumn,amsmath,amssymb,floatfix]{revtex4}

\usepackage{graphicx}
\usepackage{dcolumn}
\usepackage{bm}

\graphicspath{{Figures/}}
\setkeys{Gin}{width=\linewidth}

\begin{document}

\title{Magnetotransport in C-doped AlGaAs heterostructures}

\author{B. Grbi\'{c}$^{*}$, C. Ellenberger$^{*}$, T. Ihn$^{*}$, K.\
  Ensslin$^{*}$, D. Reuter$^{+}$, and A. D. Wieck$^{+}$}

\affiliation{$^{*}$Solid State Physics Laboratory, ETH Z\"urich,
  8093 Z\"urich, Switzerland,\\$^{+}$Angewandte Festk\"{o}rperphysik,
Ruhr-Universit\"{a}t Bochum, 44780 Bochum, Germany}
\vspace{-0.5 cm}
\begin{abstract}

High-quality C-doped p-type AlGaAs heterostructures with
mobilities exceeding 150 000 cm$^2$/Vs are investigated by
low-temperature magnetotransport experiments. We find features of
the fractional quantum Hall effect as well as a highly resolved
Shubnikov-de Haas oscillations at low magnetic fields. This allows
us to determine the densities, effective masses and mobilities of
the holes populating the spin-split subbands arising from the lack
of inversion symmetry in these structures.

\end{abstract}

\maketitle
\vspace{-0.5 cm}
Modulation doping in p-type GaAs/AlGaAs heterostructures was
introduced with Be as a dopant on (100) structures
\cite{Stormer80}. Subsequently the integer \cite{Stormer83a} and
the fractional \cite{Stormer83b} quantum Hall effects were
observed in p-type material and transistor action in such devices
was demonstrated \cite{Stormer84a}. Although the mobility could be
improved to several 10$^4$ cm$^2$/Vs \cite{Stormer84b}, the
further developement of such devices remained limited, presumably
because of the diffusion of Be dopants in the structure. The
peculiar valence-band structure  and the effects of inversion
asymmetry were analyzed in detail \cite{Eisenstein84}. A revival
of research on p-type heterostructures started with Si-doping on
(311) surfaces (for a review see Ref. \onlinecite{Davies91}).
Ultra-high mobilities exceeding 10$^6$ cm$^2$/Vs were achieved and
such samples became relevant for the investigation of the
Òmetal-insulator transitionÓ in two dimensions
\cite{Papadakis98,Hamilton01,Proskuryakov02, Rahimi03, Leturcq03}.
In a series of pioneering publications by the Princeton group
two-dimensional hole gases in GaAs were analyzed in detail by
transport experiments. This included the investigation of tunable
spin splitting \cite{Lu98}, ballistic transport \cite{Lu98,Lu99},
in-plane magnetoresistance  in view of spin polarization of the
system \cite{Noh01, Tutuc01}, a detailed analysis of the low-field
resistance in a perpendicular magnetic field \cite{Papadakis02},
as well as an understanding of the Rashba effect \cite{Winkler02}
in these systems. An overview of the bandstructure of p-type GaAs
is presented in \cite{Winkler03}.

\begin{figure}[h]
  \begin{center}
    \includegraphics{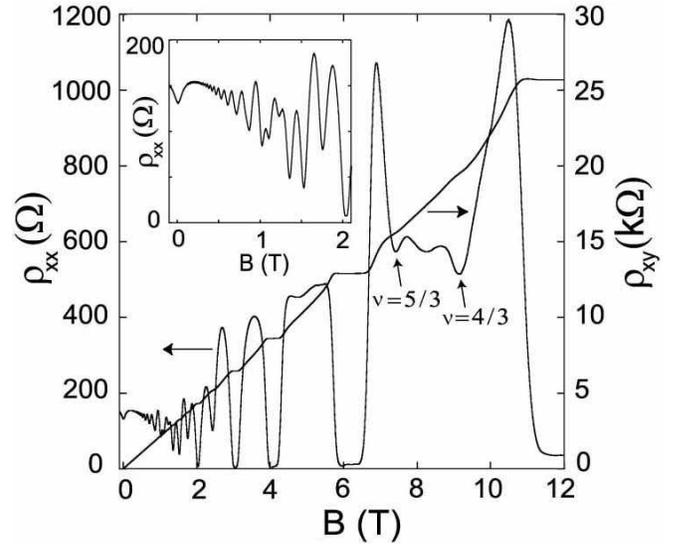}
    \caption{Magnetoresistance of a p-type  AlGaAs heterostructure. The minima related to fractional quantum Hall states are marked by vertical arrows. Inset: Blow-up of the low-field regime. A clear minimum of the magnetoresistance around B=0 as well as a beating pattern in  the Shubnikov-de Haas oscillations is observed.}
    \label{fig1}
  \end{center}
\end{figure}

Recently it has become possible to dope GaAs with C acting as an
acceptor \cite{Wieck00} on (100) substrates. Such samples have
been patterned into Hall geometries and equipped with Ohmic
contacts. Best results have been obtained for alloyed In/Zn/Au
contacts. We have investigated a series of different samples at
temperatures between 100 mK and 4.2 K. In the following we present
results obtained at 100 mK hole temperature on a sample with a
mobility of 160 000 cm$^2$/Vs and a carrier density of
3$\times$10$^{11}$ cm$^{-2}$. The two-dimensional hole gas (2DHG)
is located 100 nm below the surface. Figure \ref{fig1} shows the
magnetoresistance $\rho_{xx}$ and the Hall resistance $\rho_{xy}$.
Pronounced integer quantum Hall plateaus and minima in the
magnetoresistance at fractions of 4/3 and 5/3 document the high
quality of the sample. The inset shows the regime at low magnetic
fields. Two different oscillation periods are clearly visible. The
two periods are related to two spin-split subbands, as has been
previously shown for Be-doped (100) structures \cite{Stormer83a}
and for Si-doped (311) samples \cite{Lu98}. This spin splitting
can be attributed to the effective in-plane magnetic field
$\mathbf{B_{eff}}$, which is induced by strong spin-orbit coupling
due to structural inversion asymmetry \cite{Winkler03}. It should
be noted that $\mathbf{B_{eff}}$ varies in magnitude and
orientation as a function of in-plane wave vector $\mathbf{k}$.
Even though for a given $\mathbf{k}$ the states are spin
polarized, after averaging over all occupied $\mathbf{k}$ states
the total spin polarization in each of the two subbands vanishes.
In the following we will denote the two subbands by spin up and
spin down, having in mind the spin orientation of the states with
respect to the direction of local effective magnetic field. Due to
the different dispersion relations of the two subbands, the higher
energy spin subband is more populated with holes than the lower
energy spin subband for a given Fermi level. The difference of the
two densities, $\triangle$N, will be called carrier imbalance. For
the data presented in Fig. \ref{fig1} there are two times more
holes in the higher energy spin subband than in the lower energy
spin subband.

\begin{figure}
  \begin{center}
    \includegraphics{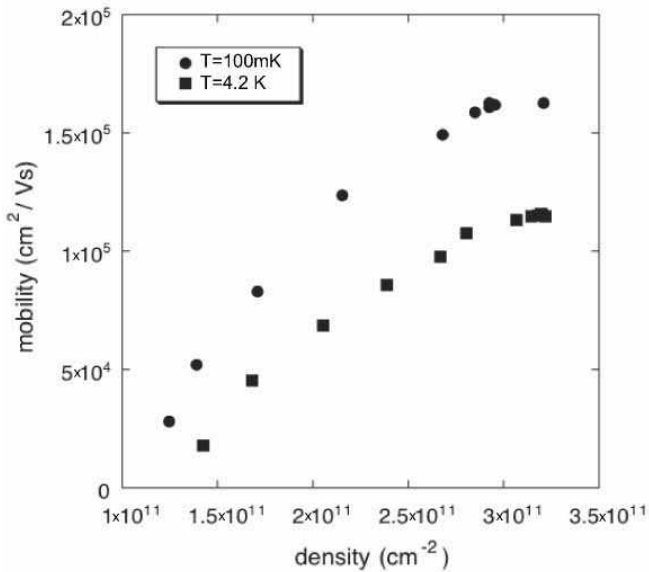}
    \caption{Mobility versus carrier density tuned with a top gate voltage.}
    \label{fig2}
  \end{center}
\end{figure}

With a homogeneous top gate (Ti/Au) the electron density and
mobility could be varied, as shown in Fig. \ref{fig2} for two
different temperatures. The average mobility is determined from
the resistivity at zero magnetic field, $\rho_{xx}$, not taking
into account the two-subband occupancy. The mobility keeps
increasing upon cooling of the sample even at temperatures below
1K possibly because of reduced phonon scattering. It is apparent,
that the mobility drops rather dramatically with decreasing
carrier density and that pinch-off is determined by a mobility
edge rather than by a vanishing carrier density.

\begin{figure}
  \begin{center}
    \includegraphics{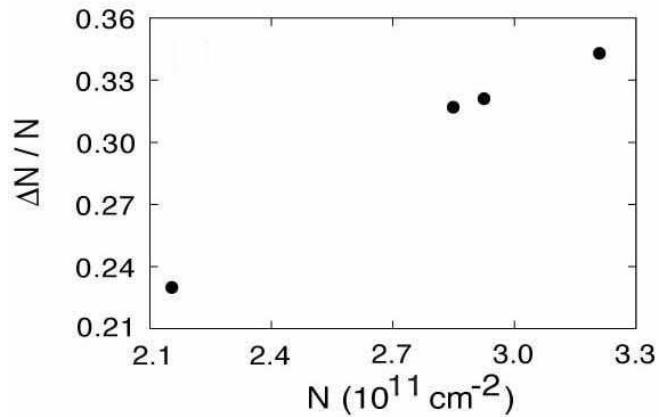}
    \caption{Ratio of the carrier densities of the two subbands versus total density for a temperature of T=100 mK.}
    \label{fig3}
  \end{center}
\end{figure}

After subtracting a second order polynomial background and
multiplying with a smoothing window, the low-field ($0.3<B<2$T)
Shubnikov-de Haas oscillations have been Fourier-analyzed and the
occupation of the two subbands was extracted. The corresponding
data is plotted in Fig. \ref{fig3} as a function of total density. The relative carrier imbalance
increases with total density as expected from bandstructure
calculations \cite{Winkler03}. As the temperature is raised from
100 mK to 1 K the carrier imbalance remains constant within measurement accuracy.

The mobilities of the individual subbands are analyzed under the
assumption that the total conductivity of the system is composed
of the conductivities of two individual subsystems. Intersubband
scattering is not taken into account. The conductivity in the
Drude model develops a parabolic B-dependence in the classical
regime \cite{Houten88,Salis99}, where Landau quantization is not
yet important.

\begin{figure}
  \begin{center}
    \includegraphics{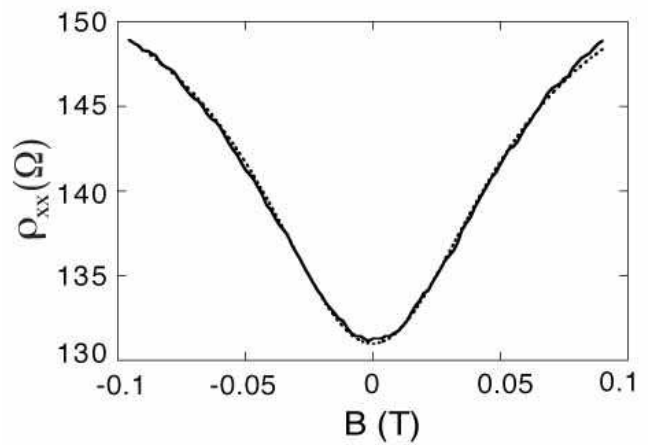}
    \caption{Experimental low-field magnetoresistance (full line) and parabolic fit (dashed line) as described in the text.}
    \label{fig4}
  \end{center}
\end{figure}

Figure \ref{fig4} shows measured low-field magnetoresistance
(solid line) together with the parabolic fit (dotted line). Taking
the carrier densities of the two subbands as extracted from a
Fourier analysis of the Shubnikov-de Haas oscillations as input
parameters (N$_s^1$= 1$\times$10$^{11}$ cm$^{-2}$ and N$_s^2$=
2$\times$10$^{11}$ cm$^{-2}$) we obtain the mobilities of the two
subbands as fitting parameters ($\mu_1$=265 000 cm$^2$/Vs and
$\mu_2$=110 000 cm$^2$/Vs).  The classical Hall effect becomes
non-linear in the two-subband model. Using the above parameters we
obtain satisfactory agreement between the measured and calculated
curve.

As it is shown in Fig. \ref{fig4}, our 2DHG exhibits a very strong
positive magnetoresistance at small perpendicular magnetic fields
($|B|\leq 0.1$T). For the highest achievable densities in our
system, 3.2$\times$10$^{11}$ cm$^{-2}$, the relative
magnetoresistance, ${(\rho_{xx} (0.1T)-\rho_{xx} (0))/\rho_{xx}
(0)}$ is 18\%. By reducing the density, at fixed temperature 100
mK, the relative magnetoresistance continuously drops and reaches
4\% at the lowest density 1.4$\times$10$^{11}$ cm$^{-2}$. Also, by
increasing the temperature from 100 mK to 1K at fixed denisty
3$\times$10$^{11}$ cm$^{-2}$, the relative magnetoresistance drops
from 17\% to 5\%. This is in agreement with previous results
\cite{Papadakis02}, and represents clear evidence that there is a
strong correlation between carrier imbalance in spin-split
subbands with different mobilities and positive magnetoresistance.

From the temperature dependence of the amplitude of the low-field
SdH oscillations the hole effective mass in the high-mobility spin
subband is determined to be $m_1=(0.34\pm0.01)$ $m_e$, which is in
agreement with previous results for the effective mass in (100)
plane \cite{Stormer83a}. Using the Boltzmann result for the
elastic scattering rate \cite{Davies} and the assumption that
carriers from both subbands see the same scattering potential
gives $m_2/m_1=(\mu_1/\mu_2)^{1/2}$. This relation, together with
the previously obtained results for $m_1, \mu_1$ and $\mu_2$,
gives $m_2$ = 0.53 $m_e$ for the effective mass of the
low-mobility spin subband.

The quantum scattering time is obtained by fitting an envelope
function to the low-field SdH oscillations,
$\tau_q=2.3\times10^{-12}$s. The zero-field longitudinal
resistance gives $\tau_D=3\times10^{-11}$s for the Drude
scattering time. The ratio $\tau_D/\tau_q = 13$ indicates that a
long range scattering potential is dominant.

By increasing the temperature from 100mK to 1K, the zero-field
resistivity of the 2DHG increases by 6\%. This behavior, which has
been interpreted as a metallic state of a 2DHG
\cite{Papadakis98,Hamilton01,Proskuryakov02, Rahimi03, Leturcq03},
is another confirmation of the sample quality.

Our results demonstrate that C-doped (100) AlGaAs-GaAs
heterostructures show reasonable electronic properties comparable
to the more established Si-doped (311) samples. Tuning of the
carrier density with a homogenous top gate has been demonstrated.
With a mean free path larger than 1$\mu$m such structures lend
themselves for the fabrication of nanostructures either with split
gates defined by electron beam lithography \cite{Daneshvar97} or
by local oxidation of the surface with an AFM
\cite{Held98,Rokhinson02}. This would allow the realization of
quantum dots and quantum point contacts with larger interaction
strength, stronger spin-orbit interaction and larger g-factor
compared to n-type GaAs heterostructures.

We would like to  thank Roland Winkler and Mansour Shayegan for valuable discussions.
Financial support from the Swiss National Science Foundation is
gratefully acknowledged.

\end{document}